\documentclass[twocolumn]{aastex701}

\usepackage{physics}
\usepackage{bm}
\newcommand{\msun}{{\rm M}_\odot}
\newcommand{\mps}{M_{\rm ps}}

\begin{document}

\title{How External Medium outside Prestellar Cores Affects Protostellar Growth: Variations in Accretion Rate and Evolution of Disks and Outflows}

\correspondingauthor{Shingo Nozaki}
\author[orcid=0000-0003-4271-4901]{Shingo Nozaki}
\affiliation{Department of Earth and Planetary Sciences, Faculty of Science, Kyushu University, Nishi-ku, Fukuoka 819-0395, Japan}
\email[show]{nozaki.shingo.307@s.kyushu-u.ac.jp}

\author[orcid=0000-0002-0963-0872]{Masahiro N. Machida} 
\affiliation{Department of Earth and Planetary Sciences, Faculty of Science, Kyushu University, Nishi-ku, Fukuoka 819-0395, Japan}
\email{machida.masahiro.018@m.kyushu-u.ac.jp}

\begin{abstract}
We investigate how the external medium surrounding prestellar cores affects the star formation process by conducting three-dimensional resistive magnetohydrodynamic simulations. The initial cores follow critical Bonnor-Ebert profiles and are embedded in environments with different ambient densities. The simulations follow the evolution at least until the envelope mass within a radius equal to twice the critical Bonnor-Ebert radius drops to $35\,\%$ of the initial cloud mass. We reveal that in environments with higher external density, enhanced mass inflow from the envelope leads to Bondi-like accretion as the protostellar mass increases. The continued inflow substantially increases the final stellar mass, resulting in star formation efficiencies that appear to exceed unity in dense environments. The external medium also influences the evolution of circumstellar disks and protostellar outflows: with the high-density external medium, disks grow rapidly but their mass becomes smaller relative to the protostellar mass, and the outflow is sustained over a long duration. However, the ratio of angular momentum removed by outflows and magnetic braking to that introduced by inflowing gas decreases with increasing external density. These results suggest that the density of the external medium regulates not only protostellar mass growth but also the inflow–outflow balance and angular momentum transport in magnetized, rotating star-forming cores.
\end{abstract}


\section{Introduction} \label{sec:intro}
The origin of the stellar initial mass function (IMF), first introduced by \citet{salpeter1955}, remains a central unsolved problems in star formation theory and is critical to understanding galaxy formation and evolution. Observations of field stars, young open clusters, and star-forming regions in the Milky Way \citep[e.g., ][]{luhman2007,bochanski2010,lodieu2012,dario2012,kirkpatrick2024} uncover a universal power-law slope at the high-mass end of the IMF. The slope and overall shape are broadly consistent with empirically derived forms \citep[e.g.,][]{salpeter1955, kroupa2001, kroupa2002, chabrier2003, chabrier2005}. This universality hints at an underlying mechanism governing stellar mass assembly. However, the origin of this shape and the mechanisms that regulate stellar masses are still not fully understood.

As a potential clue to the origin of the IMF, the similarity between the core mass function (CMF) and the IMF has been widely discussed. Assuming a one-to-one mapping between cores and stars, the star formation efficiency (SFE) is estimated to be $30$–$50\,\%$, based on the CMF peak near $1\, \msun$ \citep[e.g.,][]{nutter2007, enoch2008, andre2010}. However, some studies report that the CMF peak coincides with the IMF peak, suggesting that the SFE could approach or even exceed $100\,\%$ \citep[e.g.,][]{takemura2021a,takemura2021b}. These discrepancies may partly stem from uncertainties in determining the CMF peak, which is known to vary across regions. Such variation may arise from differences in core definitions or observational limitations such as blending effects in distant star-forming regions \citep[e.g., ][]{goodwin2008}. Indeed, based on a statistical model constrained by observed binary properties and the CMF, \citet{holman2013} pointed out that if a core continues to accrete while forming a star, the effective SFE may naturally exceed $100\,\%$.

Recent observational and theoretical findings underscore the importance of not only the internal structure of prestellar cores but also the characteristics of their surrounding environments. Many prestellar cores exhibit internal density profiles that are consistent with the Bonnor-Ebert profile \citep[e.g.,][]{alves2001,kandori2005}, yet observations suggest that the ambient gas density varies substantially across different clouds and star-forming regions \citep[e.g.,][]{motte2001,nielbock2012,roy2014,palmeirim2013}. Recent theoretical studies further indicate that gas accreted onto protostars may originate from beyond the spatial extent of observationally identified cores \citep[e.g.,][]{pelkonen2021,nozaki2025}. Additionally, some observations have detected signatures of mass inflow from outside the core \citep[e.g.,][]{konyves2020,Redaelli2022,tatematsu2022,hsu2025}. Thus, these findings motivate a systematic investigation into how variations in external density affect the accretion history and final stellar mass.

The gravitational collapse of prestellar cores has been extensively studied using analytical and numerical approaches. 
Early one-dimensional analyses clarified self-similar collapse solutions, with \citet{shu1977} predicting a constant mass accretion rates under idealized conditions \citep[e.g.,][]{larson1969,penston1969,shu1977,hunter1977,whitworth1985,foster1993,vorobyov2005}. Subsequent two-dimensional and three-dimensional simulations, including MHD studies, have investigated the roles of magnetic fields and rotation in core collapse and disk formation \citep[e.g.,][]{tomisaka1996, ogino1999,machida2008}. However, these studies typically assumed isolated cores with idealized initial conditions, and the contribution of the external environment to gravitational collapse remains poorly understood.

Extending these earlier studies, in our previous study \citep{nozaki2023}, we investigated the impact of the external medium on the protostellar mass evolution by performing three-dimensional hydrodynamic simulations of the prestellar cores embedded in different ambient densities. The results reveal that in environments with higher external density, continuous inflow from the envelope can significantly enhance the protostellar mass and result in an effective SFE exceeding unity. This enhancement in accretion rate was found to correlate with the transition from self-gravitational collapse to Bondi-like accretion. However, these simulations neglected the magnetic field and core rotation, assuming idealized, spherically symmetric collapse. While this simplification enabled a clean assessment of external medium effects, it excluded key physical processes essential in realistic star-forming environments.

The magnetic field and core rotation play crucial roles in star formation, particularly in the formation of circumstellar disks and the driving of protostellar outflows \citep[e.g.,][]{blandford1982, pudritz1986}. Disk-mediated accretion and mass ejection processes contribute to angular momentum transport, thereby influencing the SFE and the star formation process \citep[e.g.,][]{machida2009,machida2012, price2012}. Observations also report a wide variety of outflow morphologies even in early evolutionary stages in various star-forming regions, suggesting that outflow properties are sensitive to the surrounding core environment \citep[e.g.,][]{arce2007, hsieh2023}.

Recent clump-scale magnetohydrodynamic (MHD) simulations have begun to explore how stellar mass and SFE vary with different core environments \citep[e.g., ][]{smith2009, smullen2020, pelkonen2021}. However, due to limited resolution, many of these studies adopted simplified accretion schemes in which a fixed fraction of the infalling gas, typically $50\,\%$, is assumed to accrete onto the protostar, without directly resolving outflow driving. While this approach is computationally practical, it limits the ability to assess the true impact of feedback processes. To accurately evaluate how external medium affects the SFE under realistic physical conditions, it is essential to conduct high-resolution local simulations that resolve disk and outflow dynamics down to au-scale.

This paper aims to extend the work of \citet{nozaki2023} by incorporating the magnetic field and rotation to investigate how the external medium of the prestellar core affects the mass accretion rate onto the protostar and the SFE under more realistic star-forming conditions. To this end, we perform three-dimensional MHD simulations of prestellar core evolution, in which Bonnor-Ebert spheres are embedded in ambient medium with different external densities. We quantitatively evaluate the impact of external density on the SFE and examine how differences in ambient density influence disk formation and outflow driving. We also discuss the possibility of Bondi-like accretion in simulations that include the magnetic field and rotation.

This paper is structured as follows. Section \ref{sec:methods} describes the numerical settings and initial conditions for the prestellar core and its surrounding region. Section \ref{sec:results} presents the results of our analysis on how the external density of the ambient matter affects the protostellar mass growth, the SFE, and the evolution of disk and outflow structures. In Section \ref{sec:discussion}, we interpret these results considering the effects of the magnetic field and rotation, and discuss the role of external medium with different densities in the SFE based on previous studies. Section \ref{sec:summary} summarizes our conclusions.

\section{numerical settings and initial clouds} \label{sec:methods}
To investigate protostellar mass growth under different external densities of prestellar cores, we have performed three-dimensional resistive MHD simulations using a nested grid code with a sink method \citep{basu2024}. The numerical setup follows \citet{nozaki2023}, while this study introduces the magnetic field and adopts a higher spatial resolution to investigate disk formation and outflow driving, focusing on the impact of different external environments. Except for the initial density profile and outer boundary conditions, our numerical settings are the same as those of model R02 in \cite{basu2024}. 

We first construct a sphere with a critical Bonnor-Ebert density profile adopting a central number density of $n_{\rm c}=6.0 \times 10^{5}\,{\rm cm^{-3}}$ and a temperature of $T_0=10\,{\rm K}$, for which the critical Bonnor-Ebert radius corresponds to $R_{\rm 1BE}=2.9 \times 10^{-2}\,{\rm pc}$.
\begin{figure*}[htbp]
     \centering
	\includegraphics[width=\textwidth]{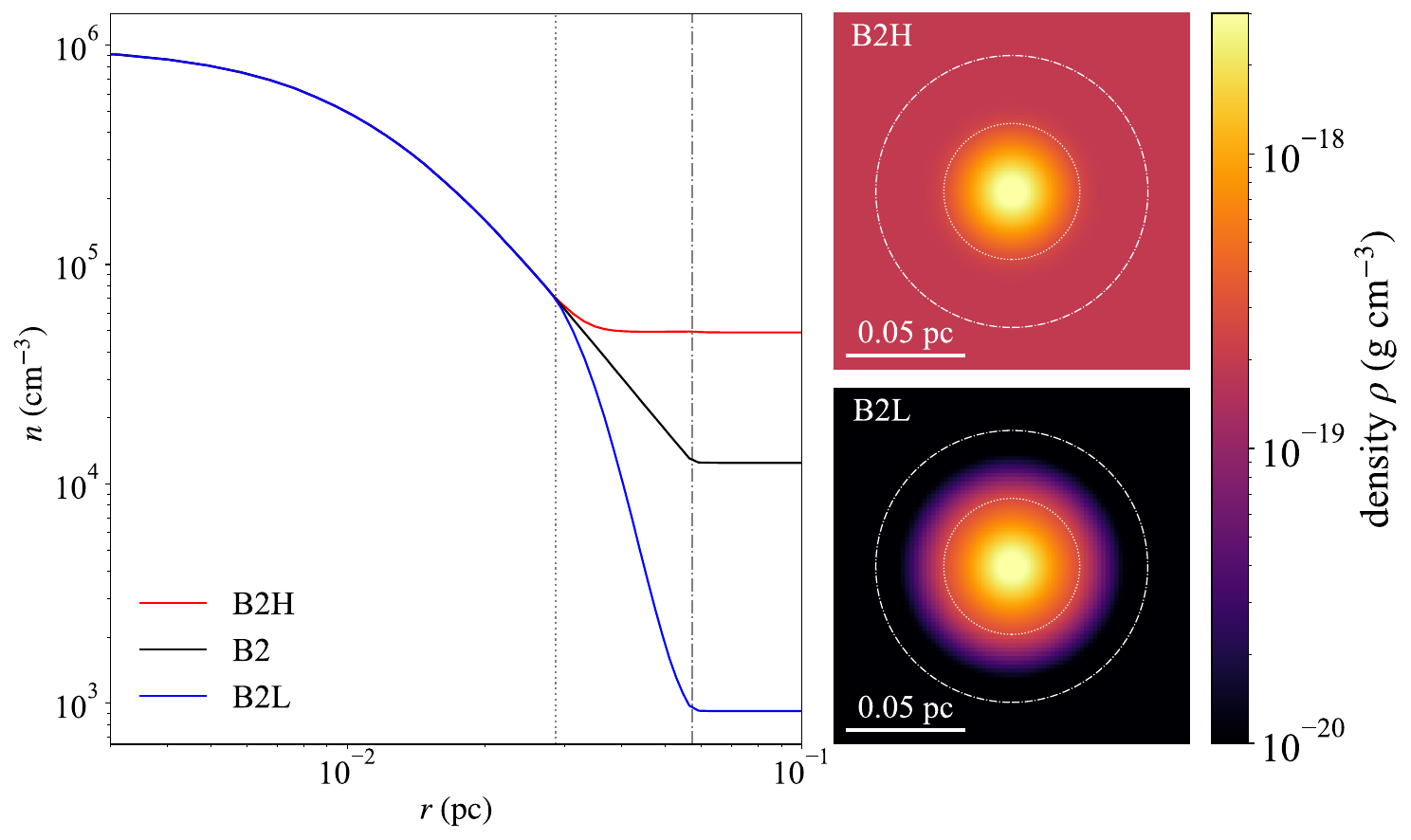} 
    \caption{The initial density profile in the radial direction for each model. 
    The gray dotted and dash-dotted lines represent one and two times the critical Bonnor-Ebert radius, corresponding to $R_{\rm 1BE} = 2.9 \times 10^{-2}\, \rm{pc}$ and $R_{\rm 2BE} = 5.7 \times 10^{-2}\, \rm{pc}$. The color maps are the 2D initial density slice in the $y=0$ plane for the model B2H and B2L. The white dotted and dash-dotted circles in the color maps indicate the positions of $r = R_{\rm 1BE}$ and $r = R_{\rm 2BE}$.}
    \label{init_BE}
\end{figure*}
To promote gravitational collapse, we then multiply the entire density profile by a factor of $f_{\rm BE}=1.68$, resulting in the central density of $n_{\rm c,0}=1.0 \times 10^{6}\,{\rm cm^{-3}}$ \citep[e.g., ][]{tokuda2020}. In the external medium region, the density smoothly decreases from the multiplied core density at $R_{\rm 1BE}$ to the external number density $n_{\rm ext}$ at $R_{\rm 2BE}$, where $R_{\rm 2BE}$ is twice the critical Bonnor-Ebert radius, following a hyperbolic tangent function. This transition ensures a continuous connection between the core and its surroundings. The specific form of the hyperbolic tangent function is given in Eq. (1) of \citet{nozaki2023}. This multiplied density profile is adopted as the initial condition in our simulations (see Table \ref{param}) and is shown in Figure~\ref{init_BE} for different $n_{\rm ext}$.

As illustrated in Figure~\ref{init_BE}, the initial cloud is defined as the combination of the prestellar core inside the $R_{\rm 1BE}$ and the external medium region between the $R_{\rm 1BE}$ and the $R_{\rm 2BE}$ \cite[see also][]{nozaki2023}. The boundary between the prestellar core and the external medium region is defined at the original critical radius $R_{\rm 1BE}$, calculated from the unscaled density profile ($n_{\rm c}=6.0 \times 10^{5}\,{\rm cm^{-3}}$, $T_0=10\,{\rm K}$). The mass enclosed within the $R_{\rm 1BE}$ in the multiplied density profile is $M_{\rm core,0}=0.98\,\msun$, which we define as the initial core mass.
\footnote{For comparison, the mass within the $R_{\rm 1BE}$ before scaling is $0.58\,\msun$, but this value is not used in this study.}
While the multiplied density profile does not completely correspond to a critical Bonnor-Ebert profile, we adopt the $R_{\rm 1BE}$ as a reference, since prestellar cores are often identified observationally as Bonnor–Ebert–like structures \citep[e.g., ][]{alves2001,kandori2005}, and we discuss this further in Section \ref{subsec:sfe}. The initial magnetic field strength is set along the $z$-axis as $B_{0}=32\,\rm \mu G$, and the initial rotation rate is $\Omega_{0}=1.4 \times 10^{-13}\,{\rm s^{-1}}$.

We consider three models, B2H, B2, and B2L, each characterized by a different external density ($n_{\rm ext}$). As listed in Table \ref{param}, the models are ordered by decreasing external density: model B2H represents a high-density environment where the surrounding gas remains dense, model B2 represents an intermediate case, and model B2L represents a low-density environment with a more diffuse surrounding medium. \citet{nozaki2023} calculated seven models with different external densities to study the impact of the environment surrounding prestellar cores on protostellar mass growth. In this study, we select three representative models, B2H, B2, and B2L, that cover the full range of external densities and capture the essential environmental variation. These models correspond to the models B2H4, B2H1, and B2L3 in \citet{nozaki2023}, with the same degree of density decrease from $R_{\rm 1BE}$ to $R_{\rm 2BE}$.

\begin{deluxetable*}{lcccccc}
\tablecaption{Models and Parameters \label{param}}
\tablewidth{0pt}
\tablehead{
\colhead{Model} & 
\colhead{$n_{\rm ext}$ (cm$^{-3}$)} & 
\colhead{$B_0$ ($\mu$G)} & 
\colhead{$\Omega_0$ (s$^{-1}$)} & 
\colhead{$M_{\rm 2BE,0}$ ($\msun$)} & 
\colhead{$\beta_0$} & 
\colhead{$\mu_0$}
}
\startdata
B2H & $4.9 \times 10^4$ & 32 & $1.44 \times 10^{-13}$ & 2.98 & 0.0525 & 3.03 \\
B2  & $1.3 \times 10^4$ & 32 & $1.44 \times 10^{-13}$ & 1.88 & 0.0299 & 2.00 \\
B2L & $9.2 \times 10^2$ & 32 & $1.44 \times 10^{-13}$ & 1.36 & 0.0271 & 1.39 \\
\enddata
\tablecomments{
Column (1): model name.
Column (2): external number density at $r = R_{\rm 2BE}$, after multiplying the initial cloud density by $f_{\rm BE} = 1.68$.
Column (3): initial magnetic field strength.
Column (4): initial rotation rate.
Column (5): initial cloud mass within $r = R_{\rm 2BE}$, after the initial cloud density is amplified by a factor of $f_{\rm BE} = 1.68$.
Column (6): initial rotational energies normalized by the magnitude of gravitational energy within $r=R_{\rm 2BE}$.
Column (7): initial mass-to-flux ratio within $r = R_{\rm 2BE}$.
}
\end{deluxetable*}

To cover a large dynamic range, we initially set six grid levels $(l = 1 - 6)$ in our nested grid code, with each grid containing $(64,\,64,\,64)$ cells. The base grid $(l=1)$ has a box size of $3.8 \times 10^5\, \rm au$, and the cell width of the $l=1$ grid is $5.9 \times 10^3\, \rm au$. The $l=1$ grid serves as the computational boundary, which is located far from the cloud surface to reduce artificial reflection of Alfv$\acute{\rm e}$n waves at the computational boundary \citep[see also][]{machida2013, machida2020b}. A finer grid is generated to ensure that the local Jeans length remains resolved with at least 16 cells \citep{basu2024}. The finest grid level $(l=13)$ has a box size of $92\,\rm{au}$, and the cell width of the $l=13$ grid is $1.44\, \rm au$.

As the outer boundary condition, we impose a gas inflow boundary at $r = R_{\rm{2BE}}$, which is twice the radius of the critical Bonnor-Ebert sphere. At this boundary (which does not correspond to the computational boundary at the $l = 1$ grid), gas inflow is prohibited, but gas outflow is allowed only if the radial velocity $v_r$ exceeds the sound speed $c_{\rm s}$ \citep[e.g.,][]{machida2009}. Beyond $r > R_{\rm{2BE}}$, both the self-gravity of the gas and the protostellar gravity are switched off. This setup ensures that the initial cloud mass remains fully conserved within a radius of $r < R_{\rm{2BE}}$, which includes the sink mass (described later) \citep{machida2013}.

To investigate long-term protostellar evolution in different envelope environments, we adopt a sink method. When the central number density on the finest grid $(l=13)$ exceeds $n_{\rm thr} = 2\times 10^{13}\, \rm cm^{-3}$, a sink cell with a radius of $r_{\rm sink} = 4\, \rm au$ is formed, which we treat as the protostar. Details on the implementation of the sink method in the nested grid code are given in \citet{machida2010a, machida2014}.

In our simulations, protostellar outflows are driven self-consistently by magnetic pressure gradients and magnetocentrifugal forces, without the use of any subgrid models. Outflows play a critical role in the star formation process and directly regulate the angular momentum and mass budget in star-forming cores. An important strength of this study is that such outflows arise naturally from the included MHD physics, rather than relying on subgrid modeling. \citet{basu2024} have also reported similar outflow driving.

Although the simulation is still running, it requires an extremely small time step, which is primarily due to high Alfv$\acute{\rm e}$n speeds and outflow velocities, and limits the simulation time. We therefore present results only up to this time, at which the envelope mass (defined in the next section) has dropped below $35\,\%$ of the initial cloud mass in all models. This indicates that the remaining $65\,\%$ has either accreted onto the protostar and disk or been ejected by outflows.

\section{Results} \label{sec:results}
\begin{figure*}[htbp]
    \centering
	\includegraphics[width=\textwidth]{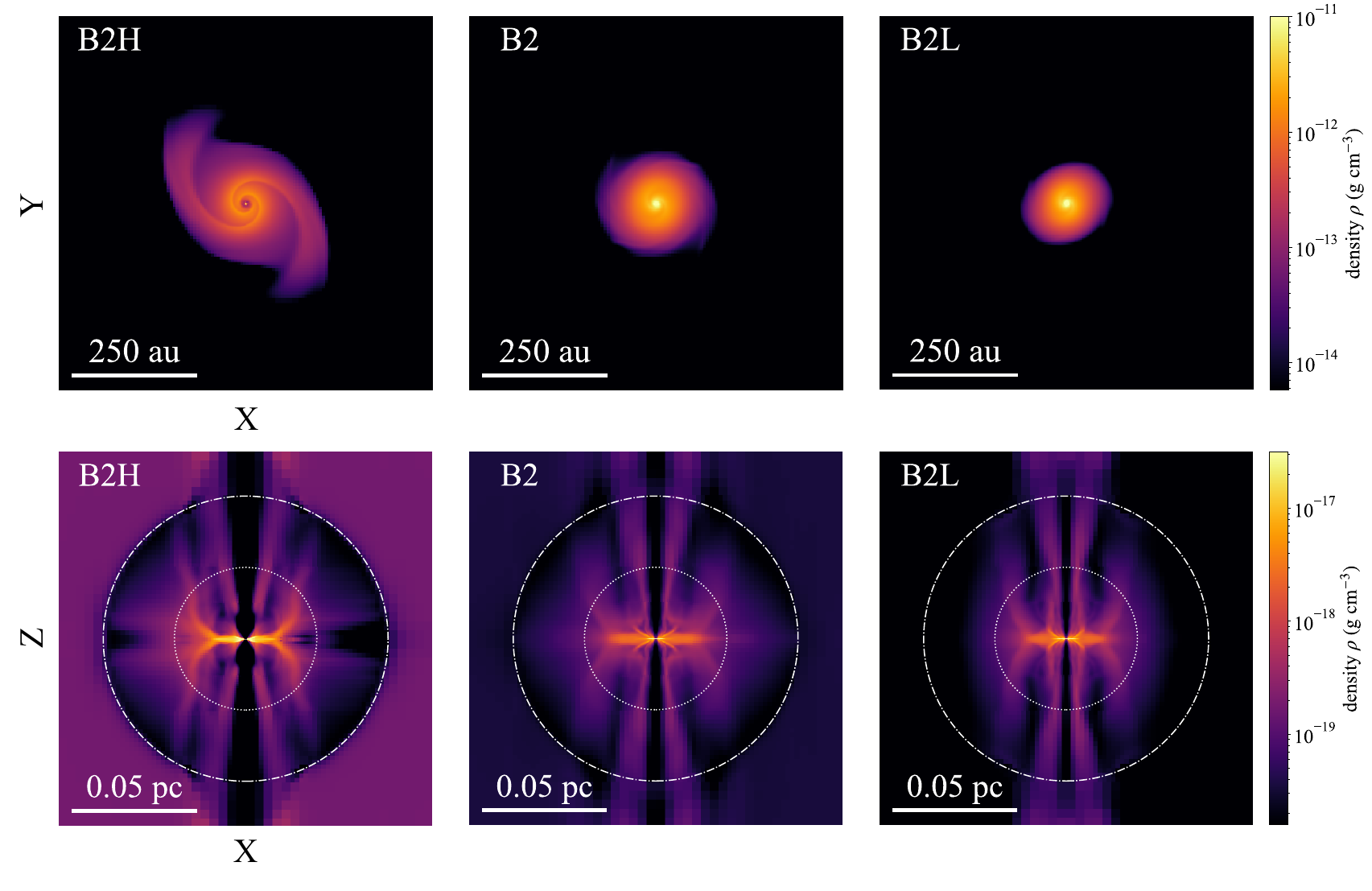}
    \caption{Density distributions on $z=0$ (top) and $y=0$ (bottom) planes for models B2H (left), B2 (middle) and B2L (right). Each panel shows the snapshot at $M_{\rm env}/M_{\rm 2BE,0}=0.35$. 
    The white dotted and dash-dotted circles in the bottom panels represent the positions of $r = R_{\rm 1BE}$ and $r = R_{\rm 2BE}$.}
    \label{snapshot}
\end{figure*}

We calculated the evolution of a prestellar core embedded in surrounding regions with different external densities. Figure~\ref{snapshot} shows the density distribution on the $z=0$ and $y=0$ planes at the moment when the envelope mass within a radius of $r=R_{\rm 2BE}$ has decreased to $35\,\%$ of the initial cloud mass. At this stage, the mass reservoir is significantly depleted, making it a suitable point for comparing differences in disk and protostellar outflow structures among the models.
In this study, the envelope mass is estimated as follows: 
\begin{equation}
    M_{\rm env} = M_{\rm 2BE,0} - \mps - M_{\rm disk} - M_{\rm out} -M_{\rm eject},
    \label{def_menv}
\end{equation}
where $M_{\rm 2BE,0}$ is the initial cloud mass within a radius of $r=R_{\rm 2BE}$, as listed in Table \ref{param}. $\mps$, $M_{\rm disk}$, $M_{\rm out}$, and $M_{\rm eject}$ represent the protostellar mass, disk mass, outflow mass, and ejected mass by protostellar outflow, respectively. The disk region is defined by the following criteria: (i) number density is above $10^8\ \mathrm{cm}^{-3}$, (ii) the azimuthal velocity $v_\phi$ in cylindrical coordinates exceeds twice the radial velocity $v_r$, and (iii) $v_\phi$ is greater than $70\,\%$ of the Keplerian velocity $v_{\rm kep}$ ($=\sqrt{G\mps/r}$). The outflow region is defined as the area where the radial velocity in spherical coordinates exceeds $50\,\%$ of the sound speed at $T = 10\rm\,K$, $v_{\rm r} > 0.5\,c_{\rm s}$. $M_{\rm eject}$ is the mass that has been expelled beyond $r = R_{\rm 2BE}$ by protostellar outflow. We have confirmed that the sum of $M_{\rm env}$, $\mps$, $M_{\rm disk}$, $M_{\rm out}$, and $M_{\rm eject}$ is always equal to $M_{\rm 2BE,0}$.

\begin{figure*}
    \centering
	\includegraphics[width=\textwidth]{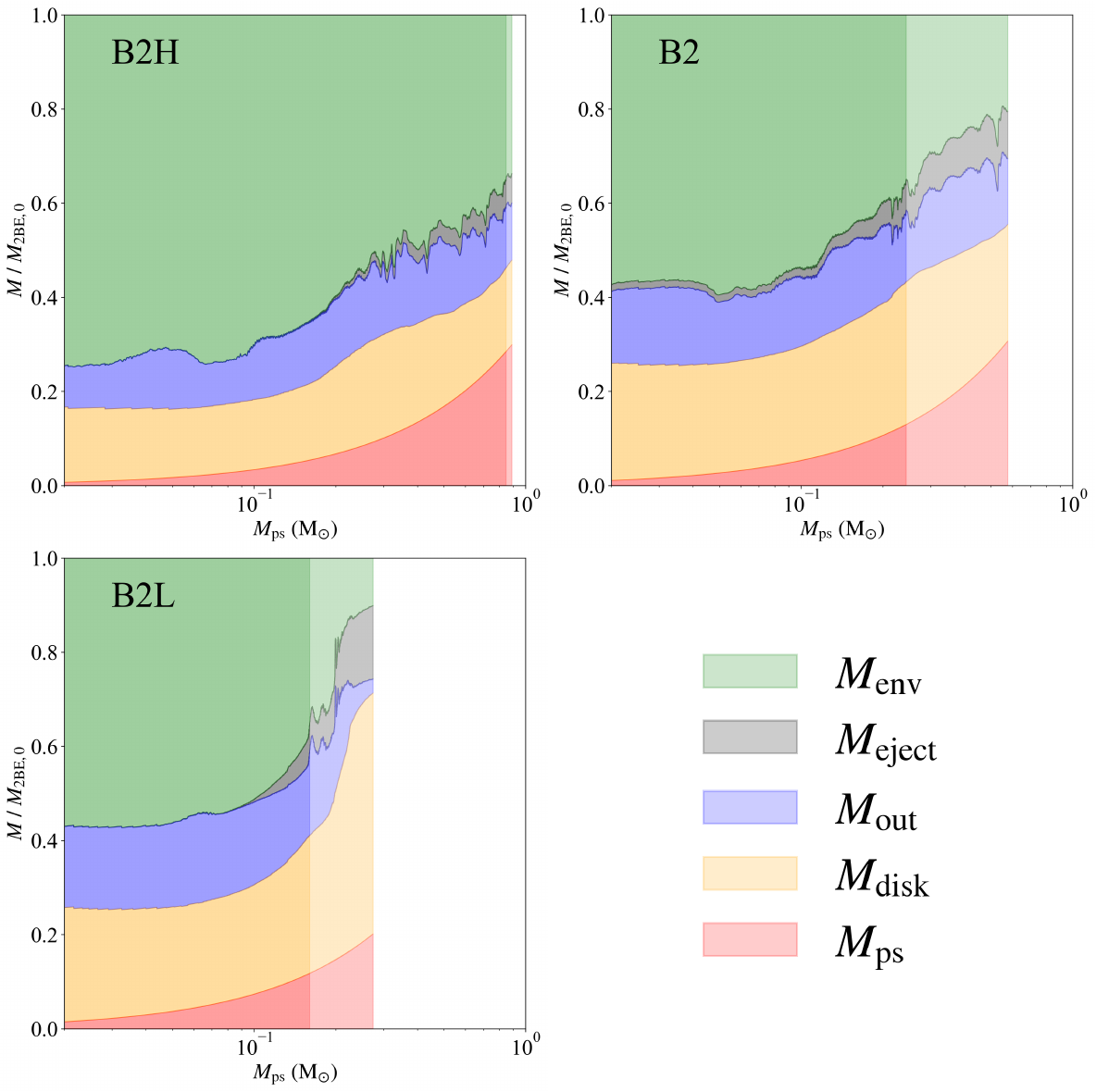}
    \caption{Mass fractions of the protostar ($\mps$, red), disk ($M_{\rm disk}$, yellow), outflow ($M_{\rm out}$, blue), ejected gas ($M_{\rm eject}$, gray), and envelope ($M_{\rm env}$, green), relative to the initial cloud mass ($M_{\rm 2BE,0}$) are plotted against protostellar mass $\mps$. The results are plotted up to the point where $M_{\rm mps} = 0.89$, $0.52$, and $0.27\,\msun$ for models B2H, B2, and B2L, corresponding to the stage each simulation has reached. For all models, the evolution after the envelope mass $M_{\rm env}$ decreases to $35\,\%$ of the initial cloud mass $M_{\rm 2BE, 0}$ is shown with lighter colors.}
    \label{mps-mratio}
\end{figure*}

In Figure~\ref{snapshot}, the top panels show that disks with different sizes are formed. Notably, the disk in model B2H shows a well-developed spiral arm structure. The bottom panels illustrate that bipolar outflows are continuously launched from the disks in all models. As a comparison of the mass distribution among components, Figure~\ref{mps-mratio} shows the mass fractions of the protostar ($\mps$), disk ($M_{\rm disk}$), outflow ($M_{\rm out}$), ejected gas ($M_{\rm eject}$), and envelope ($M_{\rm env}$) relative to the initial cloud mass ($M_{\rm 2BE, 0}$). 
In all models, as the protostellar mass grows, the envelope mass consistently decreases. Nevertheless, the evolution of the disk and outflow masses differs between models, and the relative contributions of these components vary significantly, indicating that the external environment directly affects the mass distribution in star-forming cores.

The following subsections present results focusing on how the external density surrounding the core influences the mass accretion rate onto the protostar and the SFE. We also show how the external density affects the evolution of the disk mass, the driving of outflows, and the time evolution of inflow and outflow rates at the envelope scale.

\subsection{Protostellar Mass Growth} \label{subsec:mps}
To evaluate the effect of the external medium surrounding the core on protostellar mass growth, we analyze the time evolution of the mass accretion rate onto the protostar and the protostellar mass in each model.
\begin{figure*}[htbp]
    \centering
    \begin{minipage}[t]{0.48\textwidth}
        \centering
        \includegraphics[width=\linewidth]{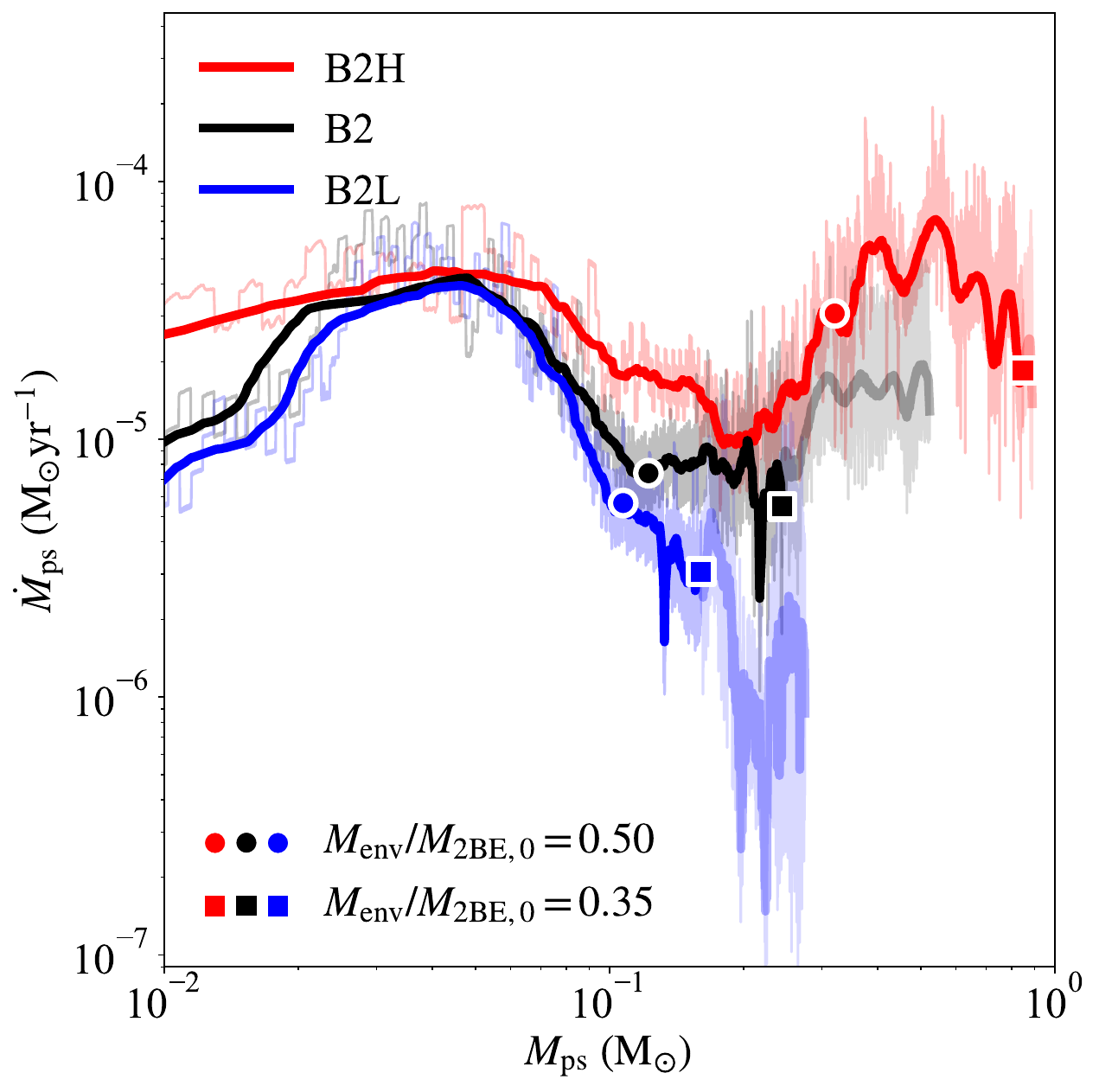}
    \end{minipage}
    \hfill
    \begin{minipage}[t]{0.48\textwidth}
        \centering
        \includegraphics[width=\linewidth]{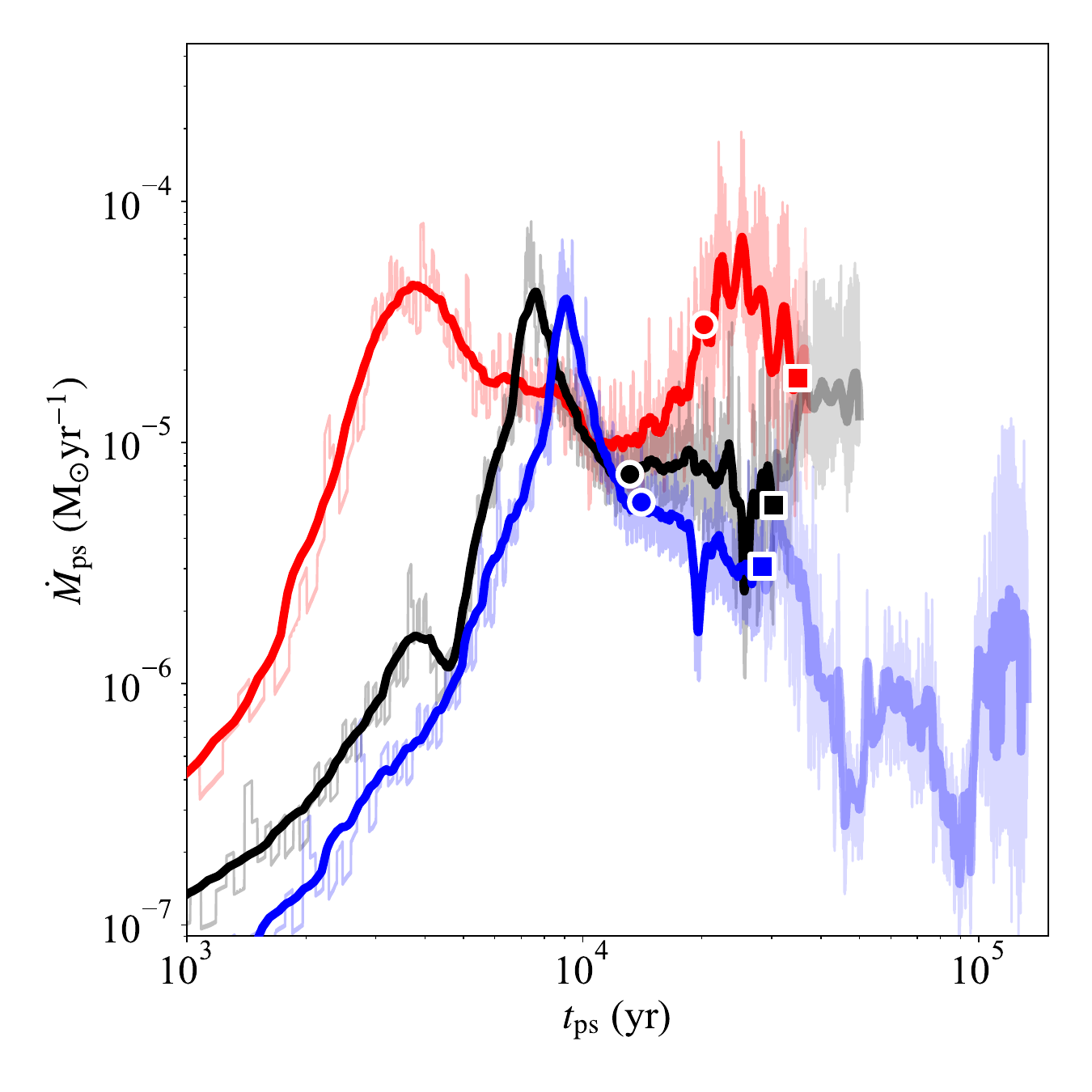}
    \end{minipage}
    \caption{Mass accretion rate onto the protostar, $\dot{M}_{\rm ps}$, for models B2H (red), B2 (black), and B2L (blue). The left panel shows $\dot{M}_{\rm ps}$ plotted against protostellar mass $\mps$, and the right panel against time since protostar formation. Thin lines indicate the mass accretion rate onto the protostars at each time. Thick lines are moving averages along the horizontal axis. The white-edged circles mark the epochs at which the envelope mass $M_{\rm env}$ has decreased to $50\,\%$ of the initial cloud mass $M_{\rm 2BE,0}$ for each model, corresponding to protostellar masses of approximately $0.32\,\msun$ (B2H), $0.12\,\msun$ (B2), and $0.11\,\msun$ (B2L). The white-edged squares mark the epochs when $M_{\rm env}$ has decreased to $35\,\%$ of $M_{\rm 2BE,0}$, corresponding to protostellar masses of approximately $0.85\,\msun$ (B2H), $0.24\,\msun$ (B2), and $0.16\,\msun$ (B2L). The subsequent evolution beyond these epochs is shown with lighter colors.}
    \label{mps-mdtps}
\end{figure*}
In Figure~\ref{mps-mdtps}, thin lines represent the mass accretion rate onto the protostar plotted against protostellar mass (left) and against time since protostar formation (right). In Figure~\ref{mps-mdtps}, the mass accretion rate onto the protostar shows short-timescale fluctuations, reflecting non-steady accretion onto the protostar.

In model B2L (blue lines in Figure~\ref{mps-mdtps}), which assumes a relatively low external density around the core, the mass accretion rate onto the protostar temporarily exceeded $\sim10^{-5}\,\msun\,\rm yr^{-1}$ before the protostellar mass $\mps$ reached $0.1\,\msun$, and then gradually declined as the protostellar mass increases. While the envelope mass decreased from $50\,\%$ to $35\,\%$ of the initial cloud mass, the protostar gained $0.05\,\msun$, corresponding to $31\,\%$ of the protostellar mass at $M_{\rm env}/M_{\rm 2BE,0} = 0.35$ in model B2L. A similar trend is found in model B2 (black lines in Figure~\ref{mps-mdtps}), which has a moderately dense external medium around the core. The mass accretion rate has a peak of $\sim 2 \times 10^{-5}\,\msun\,\rm{yr}^{-1}$ for $\mps \lesssim 0.1\,\msun$, and then gradually declined for $\mps \gtrsim 0.1\,\msun$, similar to model B2L. When $M_{\rm env}/M_{\rm 2BE,0}$ is between $0.5$ and $0.35$, the mass accretion rate is slightly higher than that in model B2L, and the protostar gained $0.12\,\msun$ during this epoch. 

In model B2H (red lines in Figure~\ref{mps-mdtps}), which is set up with a relatively high external density, the mass accretion rate onto the protostar consistently exceeded $10^{-5}\,\msun\,\rm{yr}^{-1}$ until $M_{\rm env}/M_{\rm 2BE,0}$ decreased to $0.35$. Although the mass accretion rate declined around $\mps \sim 0.1\,\msun$, it subsequently rose temporarily to a value close to $\sim 10^{-4}\,\msun\,\rm{yr}^{-1}$. This trend is very similar to that reported in \cite{nozaki2023}. When $M_{\rm env}/M_{\rm 2BE,0}$ was between $0.5$ and $0.35$, the mass accretion rate remained above $10^{-5}\,\msun\,\rm{yr}^{-1}$. During the period, the protostar gained $0.53\,\msun$, which corresponds to $62\,\%$ of the protostellar mass at $M_{\rm env}/M_{\rm 2BE,0} = 0.35$ in model B2H. While the mass accretion behavior in model B2H during $\mps = 0.01$--$0.1\,\msun$ was similar to models B2 and B2L, the subsequent mass growth phase showed a significantly different evolution.

\subsection{Comparison with Bondi Accretion} \label{subsec:bondi}
\begin{figure}[htbp]
    \centering
	\includegraphics[width=\columnwidth]{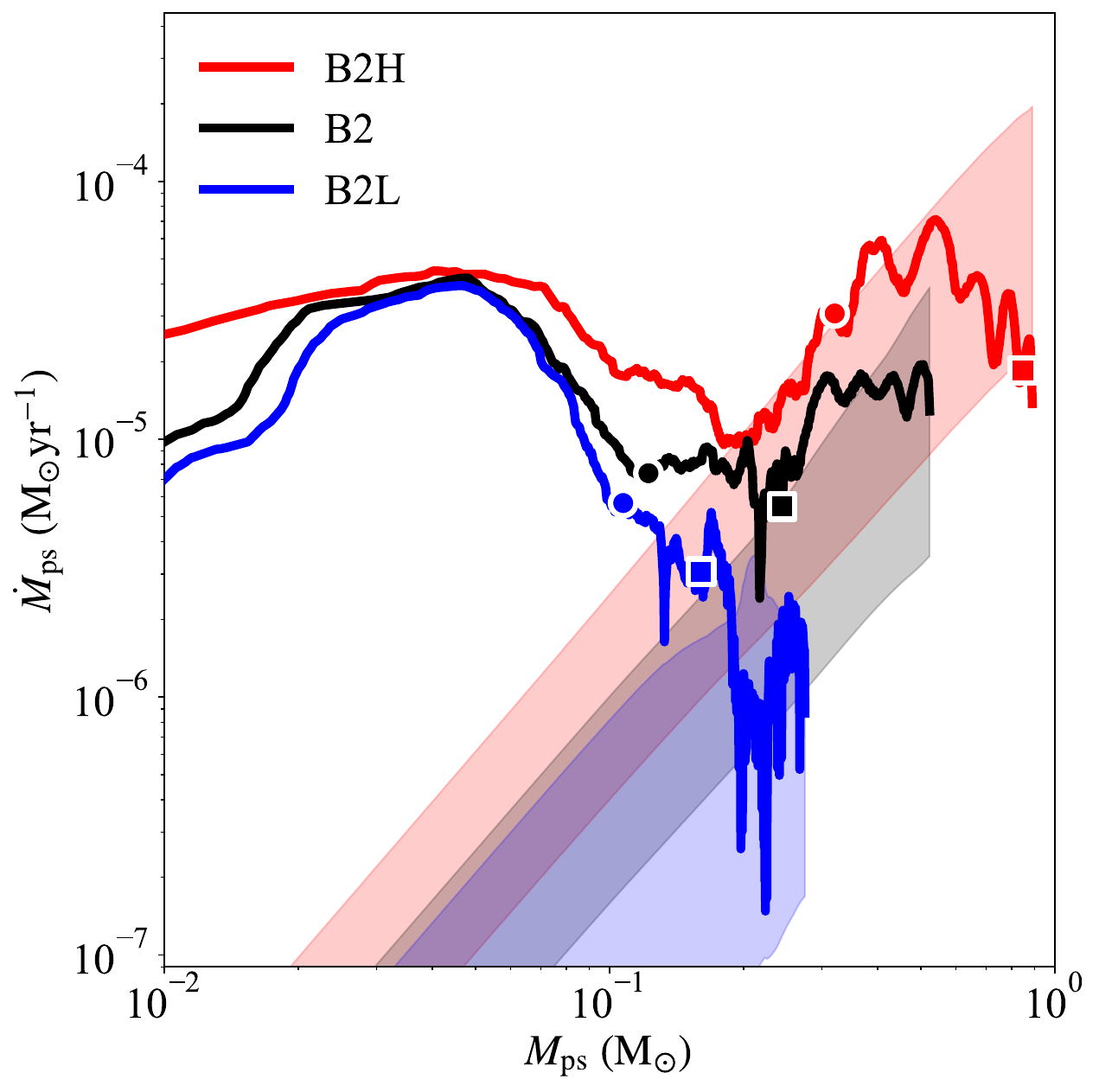} 
    \caption{Mass accretion rates onto the protostar plotted against protostellar mass for models B2H (red), B2 (black), and B2L (blue). The shaded regions indicate the range of Bondi accretion rates estimated for each model, using the gas density at $r = R_{\rm 1BE}$ (upper bound) and $r = R_{\rm 2BE}$ (lower bound).}
    \label{bondi_region}
\end{figure}

As shown in Section~\ref{subsec:mps}, the mass accretion rate onto the protostar temporarily increases in models B2H and B2. To understand this behavior, we compare the simulation results with the Bondi accretion rate calculated from the protostellar mass and surrounding gas density in the simulation, following the approach used in \citet{nozaki2023}. Figure~\ref{bondi_region} compares the expected range of Bondi accretion rates with the mass accretion rates onto protostars.
The Bondi accretion rate, which describes steady, spherical accretion onto a point mass from a uniform ambient medium \citep{bondi1952}, is given by $\dot{M}_{\rm Bondi} = 4.48 \pi G^2M^2_{\rm ps} \rho_{\infty}/c^3_{\rm s}$, where $\rho_{\infty}$ is the uniform ambient density. We used the protostellar mass at each epoch for $\mps$ and adopted the mean equatorial-plane gas density at $r = R_{\rm 1BE}$ and $r = R_{\rm 2BE}$ for $\rho_{\infty}$ ($\rho_{\infty} = \rho(r=R_{\rm 1BE})$ and $\rho_{\infty} = \rho(r=R_{\rm 2BE})$ define the upper and lower bounds) \citep[e.g., ][]{nozaki2023}. The two resulting lines define the expected range of Bondi accretion rates. 

For $\mps \lesssim 0.1\,\msun$, the mass accretion rate onto the protostar is more than an order of magnitude higher than the expected Bondi accretion rate in all models. This accretion rate onto the protostar roughly matches the value predicted by the self-similar solutions for gravitational collapse (see Section~\ref{subsec:bondidiscuss} for details). As the protostellar mass exceeds $0.2\,\msun$, the accretion rates gradually approach the Bondi accretion range. In model B2L, although no strong rise in the accretion rate is seen, the accretion rate for $\mps \gtrsim 0.2\,\msun$ remains within the expected Bondi accretion range. The Bondi accretion rate in model B2L is lower than that in the other models, reflecting the low surrounding density. In contrast, model B2H presents a clear temporary rise in the mass accretion rate onto the protostar for $\mps \gtrsim 0.2\,\msun$, closely following the upper bound of the Bondi accretion range. Model B2 also shows a modest increase in the accretion rate in the same mass range, with values largely falling within the Bondi accretion range. These results indicate that Bondi-like accretion becomes increasingly valid as the protostellar mass grows, even in the presence of the magnetic field and rotation. Notably, the transition masses at which the Bondi radius ($r_{\rm Bondi} = 2G\mps/c_{\rm s}^2$; \citealt{bondi1952}) becomes comparable to the core radius $R_{\rm 1BE}$ and the outer boundary $R_{\rm 2BE}$ are $\mps = 0.12\,\msun$ and $0.24\,\msun$, respectively. These values closely correspond to the protostellar masses at which the mass accretion rates begin to approach the Bondi accretion range, as shown in Figure~\ref{bondi_region}.

Although the Bondi radius eventually exceeds the outer boundary radius $R_{\rm 2BE}$, the gas density measured at $R_{\rm 1BE}$ and $R_{\rm 2BE}$ still serves as a reasonable proxy for $\rho_\infty$. In our simulations, the initial density profile of the core is designed to connect smoothly to the surrounding interstellar medium. Realistic prestellar cores also typically have density structures that asymptotically decline toward the ambient gas, despite some anisotropy. Thus, the densities at $R_{\rm 1BE}$ and $R_{\rm 2BE}$ can serve as reasonable approximations to $\rho_\infty$ if they represent the surrounding interstellar environment. While the Bondi formula assumes idealized conditions, we use it as a rough benchmark to interpret the accretion behavior in our simulations. A detailed comparison between the mass accretion rates onto protostars and the estimated Bondi accretion ranges is presented in Section~\ref{subsec:bondidiscuss}.

\subsection{Effects on Star Formation Efficiency} \label{subsec:sfe}
\begin{figure}[htbp]
    \centering
	\includegraphics[width=\columnwidth]{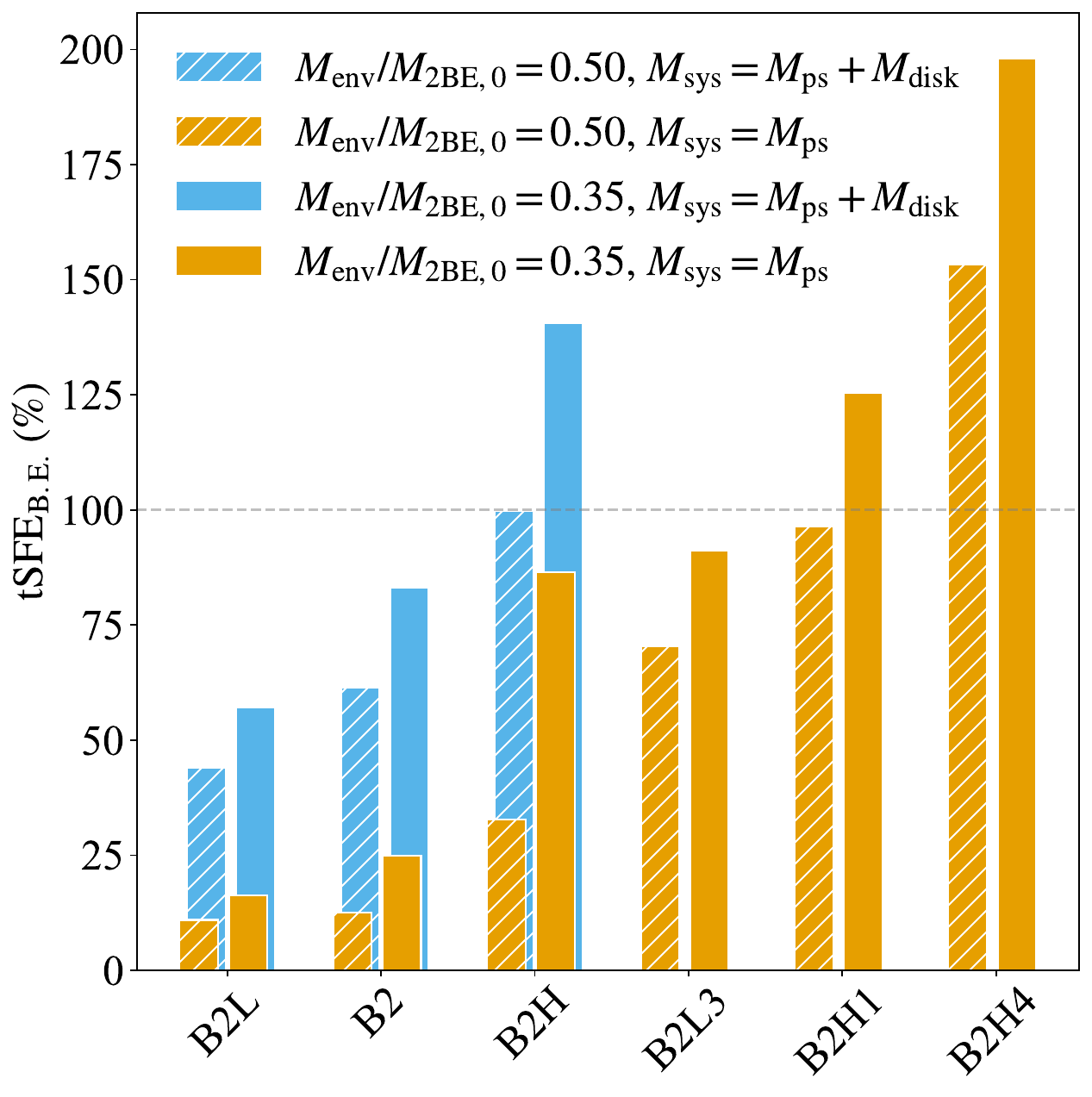} 
    \caption{$\rm tSFE_{B.E.}$ values for models B2L, B2, and B2H in this study and their corresponding non-magnetized, non-rotating counterparts from \citet{nozaki2023}. Bars with diagonal hatching indicate $\rm tSFE_{B.E.}$ values at the epoch when $M_{\rm env} / M_{\rm 2BE,0} = 0.50$, while solid bars correspond to the epoch when $M_{\rm env} / M_{\rm 2BE,0} = 0.35$. Bars in blue show $\rm tSFE_{B.E.}$ values calculated using $M_{\rm sys} = \mps + M_{\rm disk}$, and those in orange show $\rm tSFE_{B.E.}$ values using $M_{\rm sys} = \mps$.}
    \label{sfe}
\end{figure}

The SFE is typically defined as the ratio of the stellar mass to the initial mass of its natal core. In this study, following the approach of \citet{nozaki2023}, we define the time variable (apparent) star formation efficiency ($\rm tSFE_{B.E.}$) as
\begin{equation}
    {\rm tSFE_{B.E.}}(\%) = \frac{M_{\rm sys}(t)}{M_{\rm core,0}} \times 100,
\end{equation}
where $M_{\rm sys}(t)$ is either the protostellar mass $\mps$ or the sum of the protostar and disk masses $\mps + M_{\rm disk}$. The initial core mass $M_{\rm core,0}$ is calculated as the total gas mass within the Bonnor-Ebert radius $R_{\rm 1BE}$ at the beginning of the simulation, which is $0.98\,\msun$ in all models. Two evolutionary stages were selected, which correspond to epochs when the envelope mass decreased to $50\,\%$ and $35\,\%$ of its initial value. Figure~\ref{sfe} shows the $\rm tSFE_{B.E.}$ values for the three models, including a comparison with their non-magnetized, non-rotating counterparts from \citet{nozaki2023}. 
Note that, as discussed in Section~\ref{sec:methods}, we define the prestellar core as the high-density region within the $R_{\rm 1BE}$, and we therefore normalize $M_{\rm sys}(t)$ by the initial mass within the $R_{\rm 1BE}$. This definition is motivated by observations: dense cores are often identified as Bonnor–Ebert–like structures, and common core-identification methods (e.g., dendrogram analyses) may fail to recover extended, faint emission from the outer envelope, effectively leaving only the dense inner core identifiable \citep[e.g.,][]{takemura2021a, takemura2021b}. To be consistent with this observational approach, we adopt the radius $R_{\rm 1BE}$ $(=2.9 \times 10^{-2}\,{\rm pc})$ as the fiducial boundary for defining the initial core mass. Moreover, the density scaling by $f_{\rm BE}$ was applied only to promote collapse and the resulting profile does not exactly correspond to a critical Bonnor–Ebert configuration. We keep the original $R_{\rm 1BE}$ as a fiducial boundary for prestellar core. Under this definition, the protostellar mass can exceed the initial core mass. For reference, in \citet{nozaki2023}, the $\rm tSFE_{B.E.}$ was estimated by setting the envelope mass as $M_{\rm env} = M_{\rm 2BE,0} - \mps$.

At all evolutionary stages, as shown in Figure~\ref{sfe}, the $\rm tSFE_{B.E.}$ increases with higher ambient density around the core when measured using $M_{\rm sys}$. This trend is consistent with the results reported by \citet{nozaki2023}. When adopting $M_{\rm sys} = \mps$, the $\rm tSFE_{B.E.}$ at the epoch of $M_{\rm env}/M_{\rm 2BE,0} = 0.50$ is $10.9\,\%$, $12.5\,\%$, and $32.7\,\%$ for models B2H, B2, and B2L. At $M_{\rm env}/M_{\rm 2BE,0} = 0.35$, the $\rm tSFE_{B.E.}$ increases to $16.4\,\%$, $24.9\,\%$, and $86.5\,\%$ in the same models (Figure~\ref{sfe}). These values are lower than those obtained in the corresponding non-magnetized, non-rotating models (B2H4, B2H1, and B2L3), reflecting the role of circumstellar disk formation and mass ejection by outflows in magnetized and rotating environments.

When including the disk mass ($M_{\rm sys} = \mps + M_{\rm disk}$), the $\rm tSFE_{B.E.}$ at $M_{\rm env}/M_{\rm 2BE,0} = 0.50$ reaches $43.9\,\%,\, 61.5\,\%$, and $99.7\,\%$ for models B2H, B2, and B2L, and rises further to $57.1\,\%,\, 83.1\,\%,$ and $140.5\,\%$ at $M_{\rm env}/M_{\rm 2BE,0} = 0.35$. Although these values are still lower than those of their counterparts, model B2H exceeds $100\,\%$. In model B2H, the external number density is $n_{\rm ext} = 4.9 \times 10^4\,{\rm cm^{-3}}$, which corresponds to $1/20$ of the initial central number density $n_{\rm c,0} = 1.0 \times 10^6\,{\rm cm^{-3}}$. This high external density allows continuous mass accretion from outside the core, substantially enhancing protostellar mass growth and pushing the SFE above unity.

\subsection{Effects on Disk, Outflow, and Inflow} \label{subsec:disk-out-env}
The previous sections showed that the external medium surrounding the core markedly affects both the mass accretion rate onto the protostar and the SFE, even under the influence of the magnetic field and rotation. These effects are closely related to the dynamics of disk formation and outflow driving. This section presents how variations in external density affect the evolution of circumstellar disks, outflows, and mass and angular momentum transport at the envelope scale.

\subsubsection{Disk Evolution} \label{subsubsec:disk_evolution}
\begin{figure}[htbp]
    \centering
	\includegraphics[width=\columnwidth]{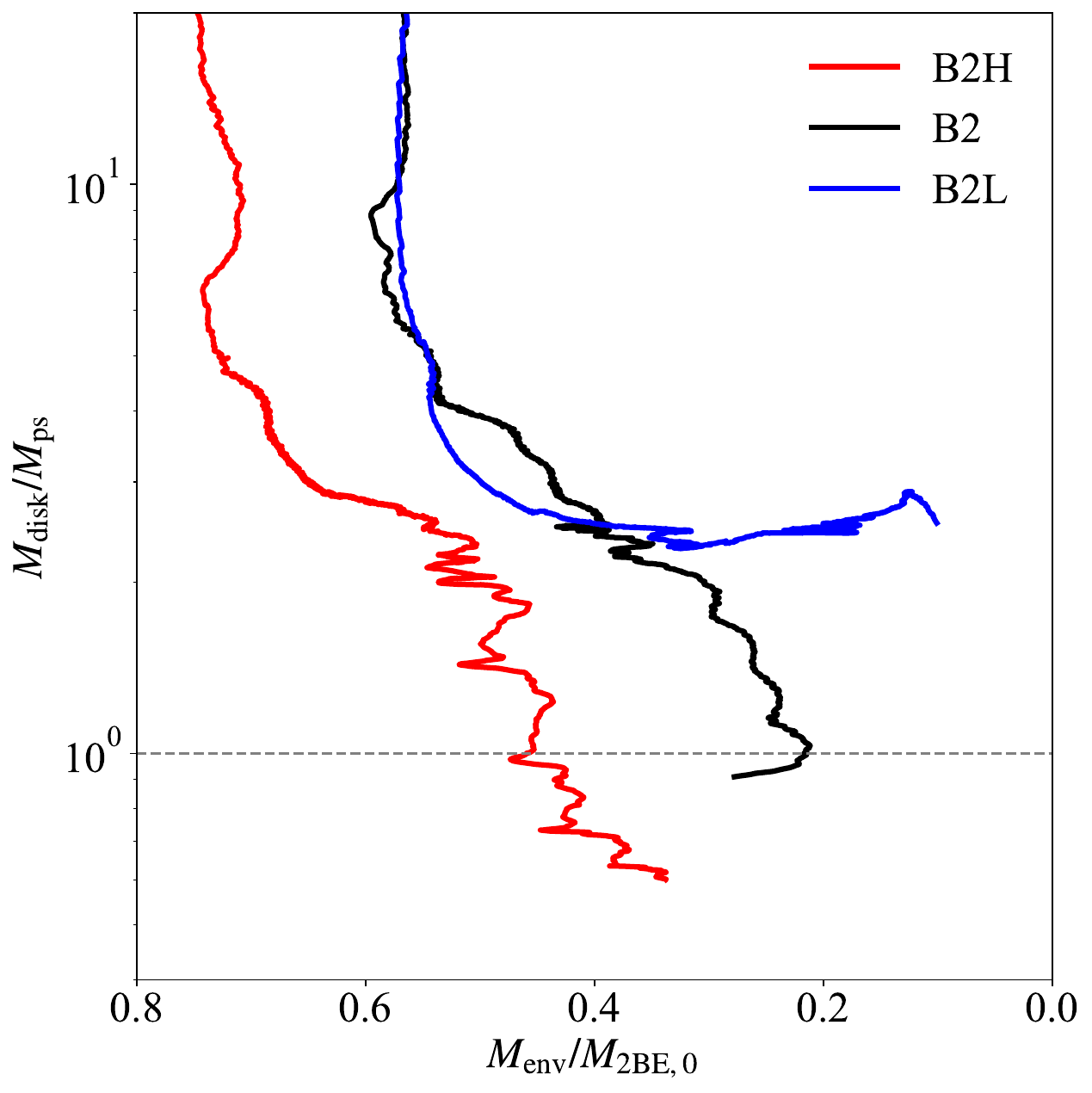} 
    \caption{The disk-to-protostar mass ratio $M_{\rm disk} / \mps$ is plotted against $M_{\rm env}/M_{\rm 2BE,0}$. Models B2H, B2, and B2L are shown in red, black, and blue lines.}
    \label{mdisk-to-mps_ratio}
\end{figure}

Circumstellar disks mediate mass transfer from the envelope to the protostar and their mass evolution sets key initial conditions for planet formation, motivating studies of how the external medium surrounding the core affect the disk-to-protostar mass ratio. As shown in Figure~\ref{mdisk-to-mps_ratio}, when a large amount of envelope mass remains (e.g., $M_{\rm env}/M_{\rm 2BE,0} \gtrsim 0.5$), the disk-to-protostar mass ratio reaches $3$ or more in all models.
This ratio in the early stages is likely related to the remnant of the first hydrostatic core, as suggested by \citet{Inutsuka2010}, which implies that the disk mass can initially exceed the protostellar mass before significant stellar growth occurs \citep[see also][]{tsukamoto2023}.

In model B2L, the disk-to-protostar mass ratio $M_{\rm disk}/\mps$ gradually decreased until $M_{\rm env}/M_{\rm 2BE,0}$ reached $\sim 0.4$. However, beyond this point, the ratio remained between 2 and 3 until $M_{\rm env}/M_{\rm 2BE,0}$ dropped to $0.1$. This result indicates that, in model B2L, the disk mass consistently exceeds the protostellar mass. Figure~\ref{mps-mdtps} shows that, for model B2L, the mass accretion rate from the envelope onto the protostar via the disk decreases, which likely contributes to maintaining the high disk-to-protostar mass ratio for $M_{\rm env}/M_{\rm 2BE,0} \gtrsim 0.4$ (for details, see below).

In model B2, $M_{\rm disk}/\mps$ decreased to around $2$–$3$ at $M_{\rm env}/M_{\rm 2BE,0} \sim 0.4$, similar to model B2L. However, as $M_{\rm env}/M_{\rm 2BE,0}$ approached $0.2$, the ratio declined further and eventually dropped below unity. In model B2H, $M_{\rm disk}/\mps$ also gradually decreased up to $M_{\rm env}/M_{\rm 2BE,0} \sim 0.4$ and is already below unity at that point. Notably, model B2H has spiral arm structures at this stage (see Figure~\ref{snapshot}), indicating enhanced angular momentum transport, which drives mass transfer from the disk to the protostar. The decline in $M_{\rm disk}/\mps$, together with a temporarily high accretion rate (see Figure~\ref{mps-mdtps}), implies that protostellar mass growth proceeds more rapidly than disk growth, as a result of this efficient transfer. As the disk-to-protostar mass ratio decreases with increasing external density, these results suggest that higher-density environments promote more efficient mass accretion onto the protostar through the disk.

\subsubsection{Outflow Evolution} \label{subsubsec:outflow_evolution} 
\begin{figure}[htbp]
    \centering
    \includegraphics[width=\columnwidth]{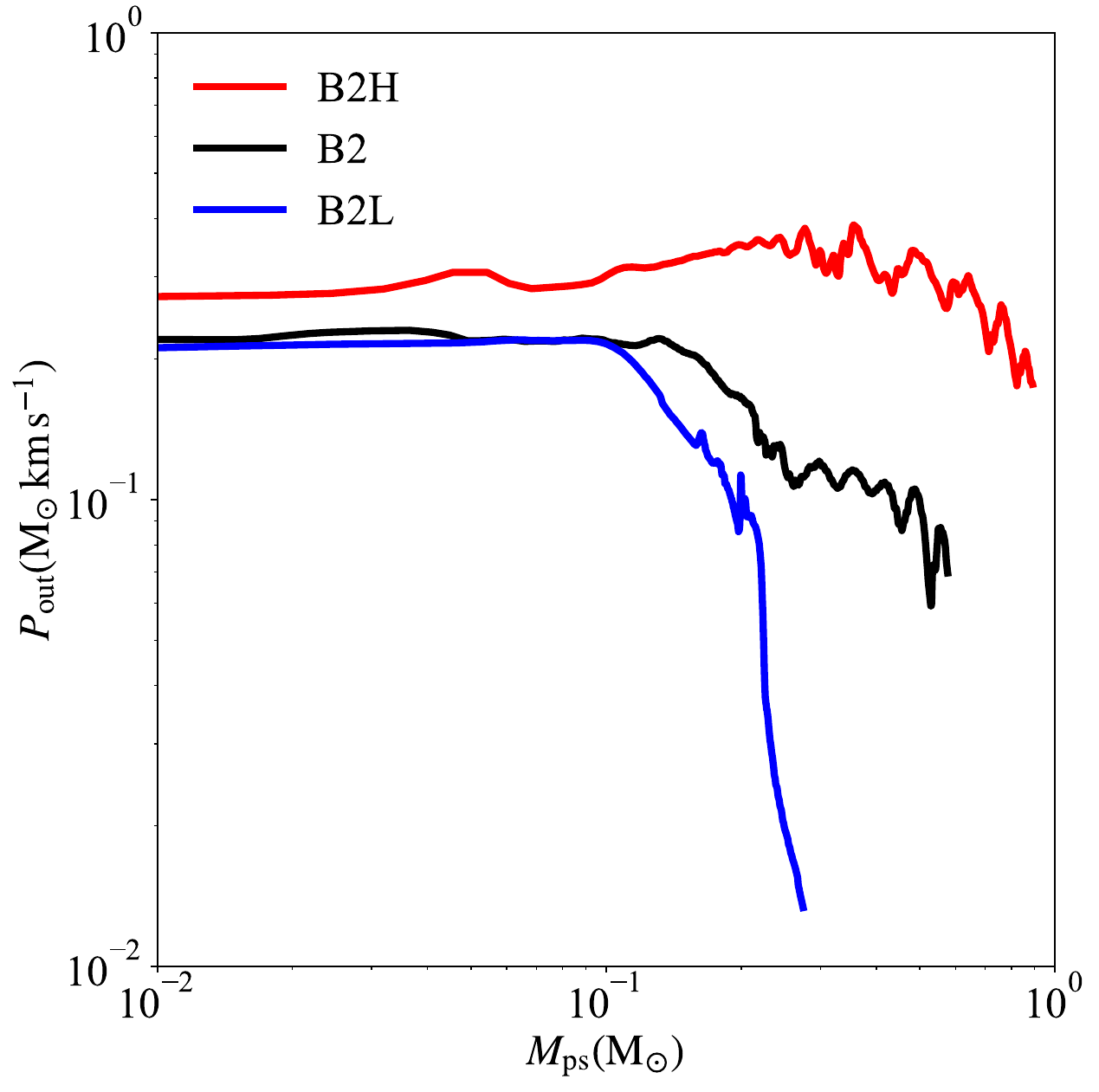}
    \caption{The outflow momentum $P_{\rm out}$ is plotted against $\mps$. Models B2H, B2, and B2L are shown in red, black, and blue lines.}
    \label{t-mvout}
\end{figure}

To investigate how the external medium around the core affects the time evolution of outflows, we analyze the outflow momentum in different models. The momentum of the outflow is particularly important, as it reflects the cumulative impact of outflow-driving processes and the energy input into the envelope \citep[e.g.,][]{matzner2000,arce2007}. Following the outflow definition described in Section~\ref{sec:results}, we evaluate the outflow momentum using
\begin{equation}
    P_{\rm out} = \int_{v_r > 0.5\,c_{\rm s}} \rho\, |\bm{v}|\, \mathrm{d}V,
\end{equation}
where $c_{\rm s}$ is the isothermal sound speed. The result is presented in Figure~\ref{t-mvout}, where the vertical axis shows the outflow momentum and the horizontal axis represents the protostellar mass.

For $\mps \lesssim 0.1\,\msun$, all models maintain nearly constant outflow momentum, with $P_{\rm out} \sim 0.2$--$0.3\,\msun\,\mathrm{km\,s^{-1}}$. For $\mps \gtrsim 0.1\,\msun$, however, the momentum in model B2L decreases rapidly by nearly an order of magnitude. In contrast, models with initially higher external density maintain higher outflow momentum. In model B2H, $P_{\rm out}$ remains nearly constant, within a factor of unity, during the increase of $\mps$ up to approximately $0.9\,\msun$. During the same period, the mass accretion onto the protostar shows a temporary increase (see the left panel of Figure~\ref{mps-mdtps}). These indicate that gravitational energy released by continuous accretion from the dense ambient environment sustains outflow driving for a longer duration.

\subsubsection{Evolution of the Envelope-Scale Mass Inflow-to-Outflow Rate Ratio} \label{subsubsec:inout_mratio} 
\begin{figure}[htbp]
    \centering
    \includegraphics[width=\columnwidth]{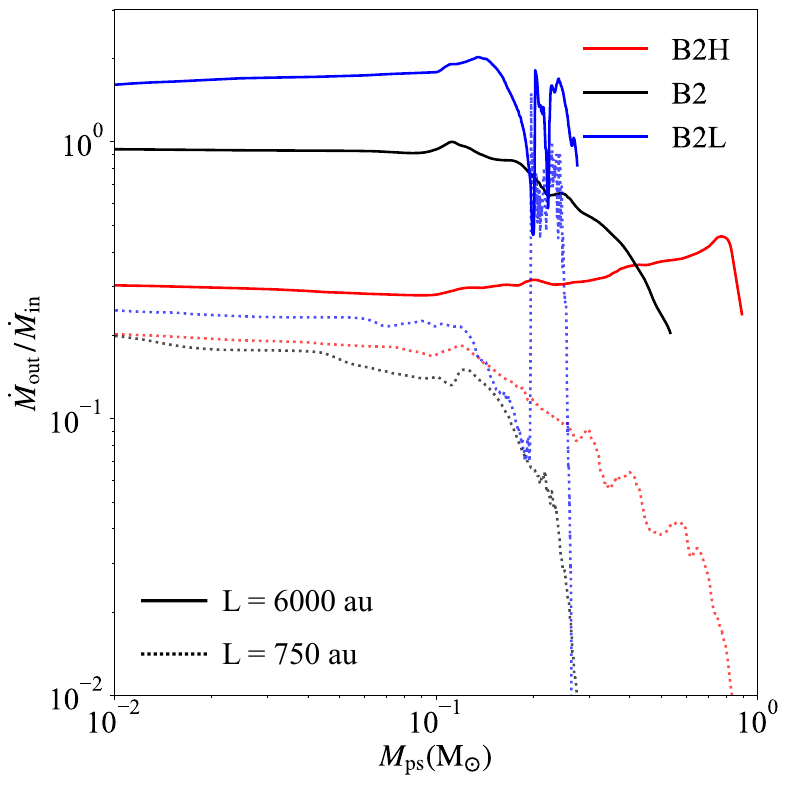}
    \caption{Ratio of mass outflow rate to inflow rate, $\dot{M}_{\rm out} / \dot{M}_{\rm in}$, plotted against the protostellar mass $\mps$. Solid and dotted lines represent box side lengths of $L = 6000$ au and $L = 750$ au. Red, black, and blue lines correspond to models B2H, B2, and B2L.}
    \label{tps-mdtinout}
\end{figure}

We analyze how the envelope-scale mass outflow-to-inflow rate ratio around the protostar varies with the external density of the core. The mass inflow and outflow rates at the envelope scale are calculated by integrating the mass flux across the six faces of a cubic box with side length $L$, centered on the protostar:
\begin{align}
    \dot{M}_{\rm in} &= \left|\int_{\bm{v} \cdot d\bm{S} < 0} \rho \bm{v} \cdot d\bm{S} \right|, \\
    \dot{M}_{\rm out} &= \left|\int_{\bm{v} \cdot d\bm{S} > 0} \rho \bm{v} \cdot d\bm{S} \right| ,
\end{align}
where $\bm{v} \cdot d\bm{S}$ is the normal component of the velocity integrated over each surface element. Figure~\ref{tps-mdtinout} presents the ratio of mass outflow rate to mass inflow rate $\dot{M}_{\rm out} / \dot{M}_{\rm in}$ plotted against the protostellar mass $\mps$ for each model. 
For $\mps \lesssim 0.1\,\msun$, the ratio $\dot{M}_{\rm out} / \dot{M}_{\rm in}$ is smaller in models with higher external density around the core than in those with lower external density, regardless of the difference between the large-scale ($L=6000\,\rm au$, solid lines) and small-scale ($L=750\,\rm au$, dotted lines) envelope regions, where the large-scale envelope region ($L=6000\,\rm au$) is located within the diameter ($2R_{\rm 1BE} \approx 12000\,\rm au$) of the critical Bonnor–Ebert sphere, i.e., inside the boundary between the prestellar core and the external medium region. Note, however, that at $L = 750\,\rm au$, model B2H shows a slightly higher $\dot{M}_{\rm out} / \dot{M}_{\rm in}$ ratio than model B2, indicating a deviation from the overall trend. These results suggest that the ratio of mass outflow to inflow is strongly influenced by the density surrounding the prestellar core. \citet{machida2020b} showed that in dense cores with initially weak magnetic fields, outflow growth can be suppressed by the high ram pressure of rapidly infalling gas in the envelope. Although the initial magnetic field in this study is not particularly weak, a similar trend is seen, likely due to the ram pressure caused by the accreting dense matter surrounding the core. In summary, the outflow rate tends to be suppressed by a high-density external medium (see also Figure~\ref{t-mvout}).

For $\mps \lesssim 0.1\,\msun$, the ratio $\dot{M}_{\rm out}/\dot{M}_{\rm in}$ shows a difference between the large-scale ($L=6000\,\rm au$) and small-scale ($L=750\,\rm au$) envelope regions. In the small-scale region, the ratio remains relatively low, in the range of 0.1–0.3. In contrast, in the large-scale region the ratio is consistently higher than that in the small-scale region across all models, and in particular, it exceeds unity in model B2L. This indicates that, for model B2L, the outflow dominates over the inflow on the large scale. Such an apparent excess is likely due to the wide-angle outflow entraining the gas of the infalling envelope \citep{matsushita2018}. For model B2H, by contrast, the ratio remains around 0.3 even in the large-scale region, reflecting the suppression of outflow growth by the dense external medium \citep{machida2020b}.

At the larger scale ($L = 6000\,\rm au$), the ratio $\dot{M}_{\rm out} / \dot{M}_{\rm in}$ begins to decline for $\mps \gtrsim 0.1\,\msun$ in models B2L and B2, and for $\mps \gtrsim 0.7\,\msun$ in model B2H. At the smaller scale ($L = 750\,\rm au$), the ratio starts to decline at $\mps = 0.1\,\msun$ in all models, although model B2L shows a temporary increase around $\mps \sim 0.2\,\msun$. In model B2L, the temporary increase occurs at the same time as a brief rise in the mass accretion rate (see Figure~\ref{mps-mdtps}), suggesting that the envelope mass was rapidly dropped, reducing the ram pressure and temporarily reactivating the outflow, as seen in \citet{machida2024}.

\subsubsection{Evolution of the Envelope-Scale Angular Momentum Fluxes} \label{subsubsec:inout_amratio} 
\begin{figure*}[htbp]
    \centering
    \begin{minipage}[t]{0.48\textwidth}
        \centering
        \includegraphics[width=\linewidth]{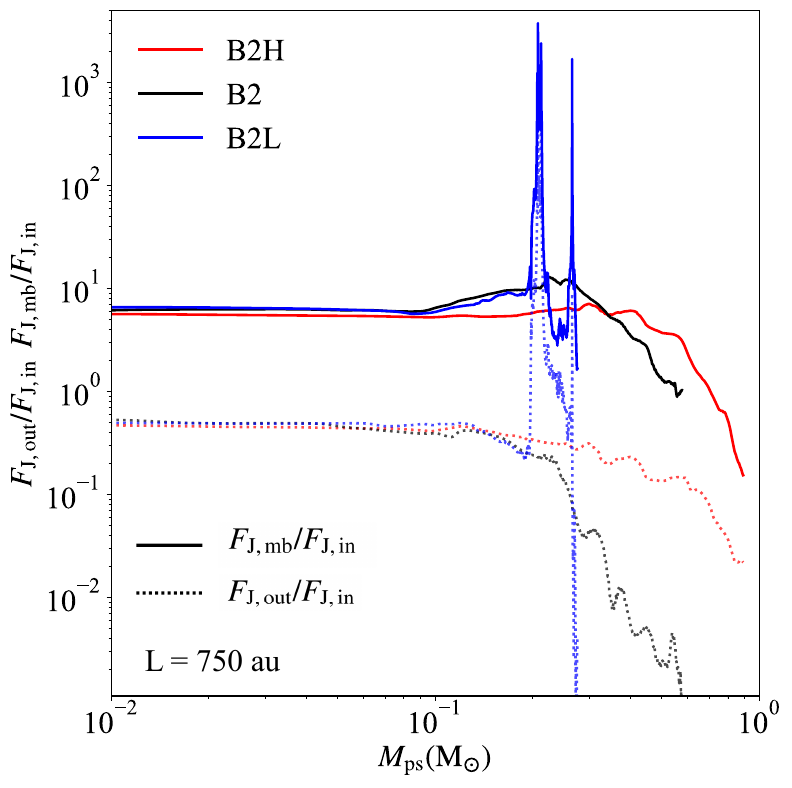}
    \end{minipage}
    \hfill
    \begin{minipage}[t]{0.48\textwidth}
        \centering
        \includegraphics[width=\linewidth]{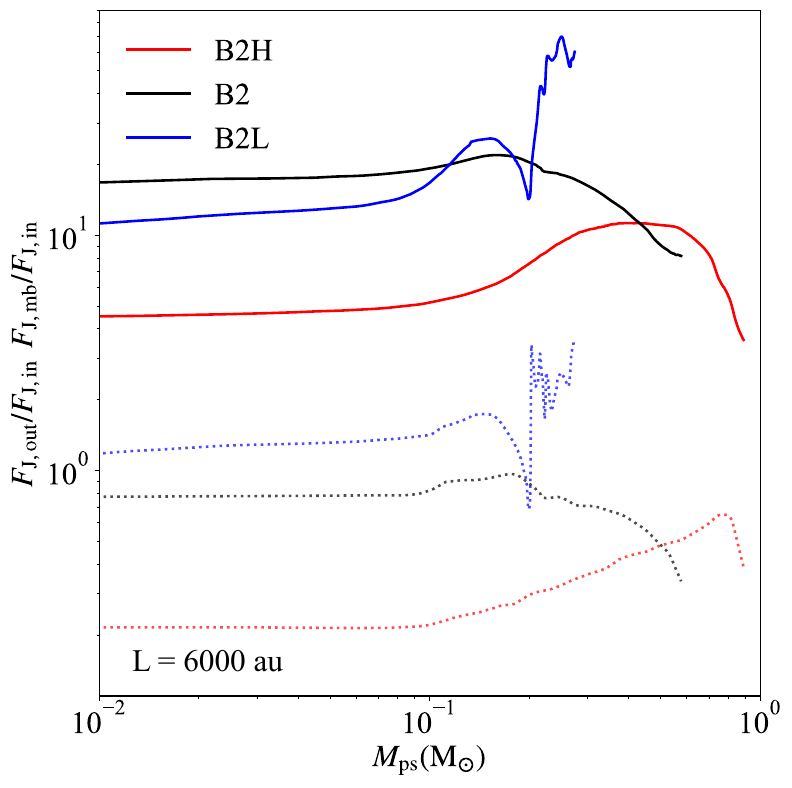}
    \end{minipage}
    \caption{Time evolution of the angular momentum flux ratios at two different envelope scales: large ($L = 6000\,\rm{au}$, right) and small ($L = 750\,\rm{au}$, left). The solid and dotted lines represent the ratios of magnetic braking and outflow angular momentum fluxes to the inflow angular momentum flux ($F_{J,\rm out}/F_{J,\rm in}$ and $F_{J,\rm mb}/F_{J,\rm in}$). Red, black and blue lines correspond to models B2H, B2 and B2L.}
    \label{tps-fj3}
\end{figure*}

To further investigate the mechanism behind the suppression of outflows in models with higher external density around the core, we estimate the angular momentum flux associated with inflow, outflow, and magnetic braking across the envelope scale. This allows us to quantify angular momentum transport under different external density conditions, which directly affects the mass accretion and ejection processes and disk growth. 
The angular momentum flux carried by the gas inflow and outflow is calculated by integrating over surfaces of cubic boxes centered on the protostar, using the following expressions:
\begin{align}
    F_{J,\rm in} = \left| \int_{\bm{v} \cdot d\bm{S} < 0} \rho r_{\rm c} v_\phi \bm{v} \cdot d\bm{S} \right|, \\
    F_{J,\rm out} = \left| \int_{\bm{v} \cdot d\bm{S} > 0} \rho r_{\rm c} v_\phi \bm{v} \cdot d\bm{S} \right|,
\end{align}
where $r_{\rm c}$ is the cylindrical radius from the rotation axis, $v_\phi$ is the azimuthal velocity, and $d\bm{S}$ is the outward-pointing surface element vector. In addition, the angular momentum flux associated with the magnetic braking is given by:
\begin{equation}
    F_{J, \rm mb} = \left| \int r_{\rm c} \frac{B_{\phi}}{4\pi} \bm{B} \cdot d\bm{S} \right|,
\end{equation}
where $B_\phi$ is the azimuthal magnetic field and $\bm{B} \cdot d\bm{S}$ is the magnetic flux through each surface element \citep[e.g., ][]{matsumoto2004,joos2012,zhao2013,machida2024}. Figure~\ref{tps-fj3} presents the time evolution of the angular momentum flux ratios at large ($L = 6000\,\rm{au}$, right panel) and small ($L = 750\,\rm{au}$, left panel) envelope scales. In this figure, the ratios of outflow and magnetic braking angular momentum fluxes are shown relative to the inflow angular momentum flux. 
These ratios provide a quantitative measure of how efficiently angular momentum is extracted from the envelope via outflows and magnetic braking.

As shown in Figure~\ref{tps-fj3}, the angular momentum flux due to magnetic braking (solid lines) exceeds that of the outflow (dotted lines) regardless of envelope scale or model. This result indicates that magnetic braking dominates outflow for transporting angular momentum in the early phase, in agreement with some previous studies \citep[e.g.,][]{joos2012,machida2020a,machida2024}. In the large scale of $6000\,\rm au$, for $\mps \lesssim 0.1\,\msun$, both $F_{J,\rm out}/F_{J,\rm in}$ and $F_{J,\rm mb}/F_{J,\rm in}$ are consistently lower in model B2H than in model B2L. This suggests that, relative to the angular momentum introduced by inflow, the removal of angular momentum by outflows and magnetic braking is less efficient in environments with higher external density around the core. In all models, at the smaller envelope scale ($L = 750\,\rm{au}$), for $\mps \gtrsim 0.1\,\msun$, $F_{J,\rm out}/F_{J,\rm in}$ drops sharply. The sharp drops suggest that, as the envelope mass decreases, outflows become inefficient in removing angular momentum. The temporary rise in $F_{J,\rm out}/F_{J,\rm in}$ and $F_{J,\rm mb}/F_{J,\rm in}$ seen in model B2L in Figure~\ref{tps-fj3}, around $\mps \sim 0.2\,\msun$, may be attributed to the same mechanism as the increase in Figure~\ref{tps-mdtinout}; a rapid decrease in envelope mass and the resulting reduction in ram pressure, which temporarily reactivates the outflow. These findings highlight the importance of external medium around the core in controlling the mass inflow–outflow balance and angular momentum transport at the envelope scale, which in turn affects the growth of stellar and disk mass in dense cores.

\section{Discussion} \label{sec:discussion}
\subsection{Bondi-Like Accretion in Magnetized and Rotating Cores}\label{subsec:bondidiscuss}
In the early phase of collapse ($\mps \lesssim 0.2\,\msun$), our simulations show that the mass accretion rate onto the protostar remains within $(0.5$--$5) \times 10^{-5}\,\msun\,\rm{yr}^{-1}$ across all models (see Figure~\ref{mps-mdtps}). This range agrees well with the self-similar solutions of gravitational collapse, which predict an accretion rate of $\dot{M} = c_{\rm acc}\,c_{\rm s,0}^3/G \sim 2 \times 10^{-6} c_{\rm acc}\,\msun\,\rm{yr}^{-1}$, where $c_{\rm acc}$ ranges from 0.97 to 46.9 \citep[e.g.,][]{larson1969,shu1977,hunter1977}. 
Adopting this range of $c_{\rm acc}$, the expected accretion rate lies between $\dot{M} = (0.2$--$7.4) \times 10^{-5}\,\msun\,\rm{yr}^{-1}$ and is in good agreement with our simulation results. For further discussion of this early accretion phase, see the discussion section of \citet{nozaki2023}.

Our simulations reveal that once the protostellar mass exceeds approximately $0.2\,\msun$, the mass accretion rate onto the protostar begins to approach the Bondi accretion rate, despite the presence of the magnetic field and rotation. Model B2H shows notable agreement between the mass accretion rate and the upper bound of the estimated Bondi accretion range when $\mps \gtrsim 0.2\,\msun$, as shown in Figure~\ref{bondi_region}. This trend persists even in the presence of disk formation and magnetically driven outflows, suggesting a transition to Bondi-like accretion \citep[see also][]{nozaki2023}.

This behavior aligns with clump-scale MHD simulations indicating continued mass inflow even after core collapse \citep[e.g., ][]{pelkonen2021}. \citet{padoan2025} has shown that mass inflow from outside the core can significantly contribute to disk mass and angular momentum. \citet{kuffmeier2024} also reveals that late-time infall can strongly affect disk orientation, especially in higher-mass systems. Our results not only support these findings but also provide the first direct evidence, at the core scale with au-scale resolution and self-consistent outflow modeling, that Bondi-like accretion emerges even under magnetized, rotating conditions.

While the mass accretion rates onto protostars closely fall within the estimated Bondi accretion range, Figure~\ref{bondi_region} shows that they slightly exceed the upper bound of the Bondi range across all models at certain evolutionary stages. This modest excess likely arises from the exclusion of disk mass in our Bondi rate estimates; incorporating disk accretion would plausibly raise the expected Bondi range. Additionally, the Bondi solution assumes a static, infinite, uniform medium \citep{bondi1952}, whereas our simulations begin with a strongly bound, collapsing core. This leads to nonzero inflow velocities and enhanced central densities, naturally resulting in some deviation from ideal Bondi accretion. Furthermore, in the later stages of Model B2H, the Bondi radius becomes larger than the simulation domain ($r_{\rm Bondi} \gg R_{\rm 2BE}$), leading to a larger discrepancy between the mass accretion rates and the Bondi accretion rate (see Figure~\ref{bondi_region}). These deviations reflect the breakdown of ideal Bondi assumptions in a finite computational domain and highlight the need for caution when interpreting or estimating Bondi accretion rates under such conditions. Note that, in star-forming regions, dense cores are typically clustered and each has its own gravitational sphere of influence. Thus, gas accretion is likely limited to a finite volume in reality as well, much like in our simulations.

Notably, several other physical quantities undergo significant changes around the same mass threshold ($\mps \gtrsim 0.1\,\msun$) at which Bondi-like accretion becomes prominent. The outflow momentum (Figure~\ref{t-mvout}), the mass outflow-to-inflow ratio ($\dot{M}_{\rm out} / \dot{M}_{\rm in}$) (Figure~\ref{tps-mdtinout}), and the angular momentum flux ratio (Figure~\ref{tps-fj3}) all show substantial changes once $\mps \gtrsim 0.1\,\msun$. These trends suggest that Bondi-like mass inflow influences both protostellar mass growth and internal feedback processes. Our results strongly indicate that continued mass supply from outside the core may play a crucial role in controlling the balance between inflow and outflow, as well as the evolution of angular momentum and the disk.

\subsection{The Impact of the External Medium Surrounding the Core on Star Formation}\label{subsec:sfediscuss}
In this study, we define the SFE as the ratio of the stellar system mass at a given time (either $\mps$ or $\mps + M_{\rm disk}$) to the initial core mass, which we estimate as the mass within the Bonnor-Ebert radius ($R_{\rm 1BE}$). We use the envelope mass at the epoch when $M_{\rm env}/M_{\rm 2BE,0} = 0.50$ or $0.35$ to calculate the SFE, mimicking observational core definitions. Compared to our previous work \citep{nozaki2023}, these simulations include both the magnetic field and rotation, offering a more realistic setup for protostellar collapse and enabling direct comparison with observations. This allows us to assess whether the external-density dependence of SFE persists under these improved physical conditions.

We found a clear positive correlation between external core density and SFE. In models B2L and B2, representing relatively isolated environments, the SFE, defined as $(\mps + M_{\rm disk}) / M_{\rm core}$, ranges from $44\,\%$ to $83\,\%$ when both stellar and disk mass are included. These values are broadly consistent with previous observational and numerical studies, which typically estimate SFE values of $\sim 30$ – $50\,\%$ in star-forming regions \citep[e.g., ][]{andre2010,machida2012,price2012}. However, in the higher-density model B2H, which represents cores embedded in denser surroundings, the SFE reaches $141\,\%$ by this definition and still approaches $100\,\%$ when defined as $\mps / M_{\rm core}$. This result supports theoretical predictions (e.g., \citealt{holman2013}) that sustained mass inflow from outside the prestellar core can raise the effective SFE well above unity.

Our findings suggest that such high SFE values can be achieved even in the presence of the magnetic field and rotational support. This result helps to explain recent observations by \citet{takemura2021a,takemura2021b}, where the protostellar mass seems to exceed the molecular cloud core mass. Such cases may not simply result from observational uncertainties; instead, they could indicate mass inflow from the surrounding envelope, although other effects, such as blending in distant regions, cannot be entirely ruled out. Still, caution is warranted: as discussed by \citet{pelkonen2021}, a strict one-to-one correspondence between the core mass and final stellar mass may not universally hold \citep[see also][]{nozaki2025}. Our results suggest that the final stellar mass is not solely determined by the mass within the initial core but is also influenced by the surrounding environment. This finding implies that accounting for an open and dynamic environment beyond the core is essential for a proper understanding of protostellar mass growth.

\subsection{Limitations and Prospects for Future Work}
In this study, we adopt the magnetized, rotating Bonnor-Ebert sphere embedded in uniform external media as idealised initial conditions. Compared to our previous work without the magnetic field and rotation \citep{nozaki2023}, this setup provides a more realistic basis for parametrically assessing how external medium influences the star formation process. The simplification allows for a controlled exploration of environmental effects, which can be difficult to disentangle in fully self-consistent, turbulent settings. However, real star-forming cores typically form within filamentary and hub–filament networks that introduce strong anisotropies in density and velocity \citep[e.g.,][]{andre2010,Andre2014,palmeirim2013,konyves2023}. As such, our simulations do not capture the mass accretion from filament-surrounding gas, core--core interactions, or turbulent ram pressure. Although our results may not be directly applicable to the full complexity of molecular clouds, this approach provides a useful reference point for interpreting more realistic simulations. These considerations highlight the need for future simulations that capture protostellar formation from non-spherical core structures, including anisotropic accretion and outflow launching.

Environmental regulation of inflow and outflow rates plays a critical role in setting protostellar and disk masses. Figure~\ref{tps-mdtinout} presents the ratio of mass outflow to inflow rates, $\dot{M}_{\rm out} / \dot{M}_{\rm in}$, and shows that in the early evolutionary stage when the protostellar mass is still low ($\mps \lesssim 0.1\,\msun$), this ratio varies by less than a factor of two within each spatial scale across models with different external densities. This result supports, to some extent, the validity of adopting subgrid outflow models in large-scale simulations. However, as the disk and protostellar masses increase, the inflow–outflow ratio exhibits strong time variability, indicating that such subgrid models may systematically overestimate the outflow mass in more evolved systems. This implies that clump-scale simulations should incorporate this non-steady behavior, either explicitly or via higher-resolution methods such as zoom-in techniques. Indeed, \citet{kuffmeier2019,kuffmeier2020} and \citet{padoan2025} demonstrated the feasibility of coupling parsec-scale turbulence with sub-au disk dynamics. A statistically robust set of zoom-in simulations covering a range of external densities, magnetic field strengths, and rotation rates would enable a quantitative assessment of how environmental diversity shapes the distributions of stellar and disk masses. Such an approach is essential for linking environmentally regulated Bondi-like accretion, as found in this study, to the statistical distribution of stellar masses defined by the IMF.

Recent observations suggest that disk sizes vary systematically across star-forming environments \citep[e.g.,][]{hendler2020,shoshi2025}, likely reflecting differences in local gas properties such as magnetic field strength and rotation. Outflow morphologies appear to depend on environmental conditions, including core geometry and external pressure \citep[e.g.,][]{hsieh2023}. Our simulations also reveal that the time evolution of key quantities, such as the disk-to-protostar mass ratio and the momentum carried by protostellar outflows, depends sensitively on the external gas density surrounding each core. To enable meaningful comparisons with observations, future high-resolution simulations should adopt more realistic initial conditions that sample a broader range of magnetic and rotational parameters. By combining these simulations with synthetic observations focusing on disk and envelope structures, one can directly test how environmental factors shape the evolution of protostars and disks. Such efforts will be essential for validating and refining the environmental effects identified in this paper, and for advancing a predictive, observation-informed theory of star formation in magnetized, turbulent clouds.

Furthermore, the time evolution of the disk-to-protostar mass ratio (Figure~\ref{mdisk-to-mps_ratio}) indicates that massive disks form in all models, with disk masses exceeding those inferred for Class I YSOs from dust continuum observations \citep[e.g., ][]{williams2019}. In particular, in model B2L, the ratio remains above unity throughout the simulation, implying a disk mass of the order of $0.5-0.6\,\msun$ at the end of the simulation. The difference in disk mass between our simulations and observations may arise from the systematic underestimation of disk masses in dust continuum observations, and also from the absence of dust growth in our simulations, which is expected to reduce disk sizes as well as their masses \citep{tsukamoto2023}. In addition, our simulations cover only the first $\sim 0.1\,\rm Myr$ after protostar formation (see the right panel of Figure~\ref{mps-mdtps}), when the protostellar systems may still be classified observationally as Class 0 objects. At this stage, in model B2L the mass accretion rate at the end of the simulation is $\sim 10^{-6}\,\msun\,{\rm yr^{-1}}$, implying a depletion timescale of $\sim 0.5\,\rm Myr$ for the disk. Thus, the disk mass may decrease substantially during $\sim 0.5\,\rm Myr$ following the dissipation of the infalling envelope. If episodic accretion occurs, it could further influence the timescale of disk depletion.

\section{Summary} \label{sec:summary}
We conducted three-dimensional resistive MHD simulations of protostellar collapse under different external densities, focusing on how the surrounding environment affects the star formation processes such as disk formation and mass accretion. This study extends our previous work \citep{nozaki2023} by including the magnetic field and rotation, thereby enabling more realistic comparisons with observations. Our main findings are summarized below:

\begin{enumerate}
\item In models with higher external density, the mass accretion rate onto the protostar temporarily increases once $\mps \gtrsim 0.2\,\msun$. This increase aligns with the Bondi accretion range estimated from the external density and the protostellar mass, suggesting the onset of Bondi-like accretion at this stage. Our simulations demonstrate that Bondi-like accretion arises even under magnetized and rotating conditions as the protostellar mass grows. The close match between the mass accretion rates and the estimated Bondi accretion rate supports clump-scale MHD simulations that suggest continued mass inflow after core collapse. Moreover, the point at which the Bondi radius becomes comparable to the core size (at $\mps \sim 0.1$--$0.2\,\msun$) coincides with the transition to Bondi-like accretion, providing a physical explanation for its onset. These findings imply that mass supply from outside the core plays a crucial role on protostellar mass growth.

\item We found that the star formation efficiency (SFE), defined with respect to the initial core mass enclosed within the $R_{\rm 1BE}$, increases with external core density even under magnetized and rotating conditions. In models representing relatively isolated environments, the SFE remained below unity and was broadly consistent with previous observational and theoretical studies. In contrast, the high-density model (B2H) yielded SFE values exceeding $100\,\%$, indicating that continued mass inflow can elevate the effective SFE well above unity. This result supports theoretical predictions of continued accretion beyond the initial core and is consistent with recent observations in which the stellar mass is not fully accounted for by the initial core mass. Importantly, our definition of the SFE reflects an observationally motivated core boundary and allows meaningful comparison across models with different external environments. These findings indicate that protostellar mass growth is governed not only by the initial core mass but also by the surrounding environment.

\item Our results also reveal that variations in the external medium surrounding cores substantially influence the evolution of circumstellar disks, protostellar outflows, and angular momentum transport at the envelope scale. In higher-density environments, mass accretion from the disk to the protostar is enhanced, rapidly leading to a decrease in the disk-to-protostar mass ratio. Cores embedded in such dense surroundings also maintain stronger outflow momentum for a longer time, owing to continued accretion from the dense envelope. However, in such environments, both the mass outflow-to-inflow rate ratio ($\dot{M}_{\rm out}/\dot{M}_{\rm in}$) and the angular momentum extraction efficiency ($F_{J,\rm out}/F_{J,\rm in}$, $F_{J,\rm mb}/F_{J,\rm in}$) are lower, indicating that outflows and magnetic braking are less significant compared to inflow. These trends highlight the importance of ambient density in regulating not only protostellar mass growth but also the inflow–outflow balance and disk evolution in magnetized collapsing cores.
\end{enumerate}

\begin{acknowledgments}
We thank the referee for their very useful comments and suggestions, which significantly improved this paper. This research used the computational resources of the HPCI system provided by the Cyber Science Center at Tohoku University and the Cybermedia Center at Osaka University (Project ID: hp200004, hp210004, hp220003, hp230035, hp240010, hp250007). Simulations reported in this paper were also performed by 2020, 2021, 2022, 2023, 2024 and 2025 Koubo Kadai on the Earth Simulator (NEC SX-ACE) at JAMSTEC. This work was supported in part by MEXT/JSPS KAKENHI Grant Number JP21H00046, JP21K03617 (MNM), and JST SPRING, Grant Number JPMJSP2136(SN). This work was also supported by a NAOJ ALMA Scientific Research grant (No. 2022-22B), and by a grant from the Hayakawa Satio Fund awarded by the Astronomical Society of Japan. We acknowledge the use of OpenAI's ChatGPT only as a grammar checking and editing tool to improve the clarity and readability of the manuscript.
\end{acknowledgments}

\bibliography{rf}{}
\bibliographystyle{aasjournalv7}

\end{document}